\documentclass[preprint,aps,showpacs,preprintnumbers,amsmath,amssymb,nofootinbib]{revtex4}
\usepackage{epstopdf}
\usepackage{epsfig}
\usepackage{graphicx}

\newcommand{\be}{\begin{equation}}
\newcommand{\ee}{\end{equation}}
\newcommand{\bea}{\begin{eqnarray}}
\newcommand{\eea}{\end{eqnarray}}

\usepackage{float}

\begin{document}
\title{\large Probing non-unitary neutrino mixing via long-baseline neutrino oscillation experiments based at J-PARC}

\author{Soumya C.  }
\affiliation{Institute of Physics Bhubaneswar, Sainik School POST, Sachivalaya Marg, Odisha-751005\\ Email:soumyac20@gmail.com (ORCID: 0000-0001-8714-504X) }

\begin{abstract}
This paper investigates the capability of long-baseline experiments, which  are making use of neutrinos that are coming from Japan Proton Accelerator Research Complex (J-PARC), in establishing the unitarity of active-neutrino mixing by ruling out the non-unitary mixing scheme as a function of true values of CP-violating phase $\delta_{\mathrm{CP}}$. It is found that T2HK can establish unitarity of active neutrino mixing at above 2$\sigma$ C.L. irrespective of neutrino mass hierarchy and true value of $\delta_{CP}$, if non-unitary (NU) parameter $\alpha_{21}$ is of the order of $10^{-2}$. Further, this paper is also discuss  the bound on NU parameter in 21 sector and sensitivity limit of these experiments in determining NU parameter.  It is found that the bounds on $\left(\alpha_{21}/2\right)$ are 0.028, 0.0026, 0.005 at 2$\sigma$ C.L. respectively for T2K, T2HK, and T2HKK. Moreover, it is also found that the sensitivity limit of T2HK on NU parameter is far better than that of both T2HKK and T2K.

\end{abstract}

\maketitle
\section{Introduction}
Ever since the phenomenon of flavor transition of neutrino \cite{exp-1,exp-2,exp-3,exp-4,exp-5,exp-6,exp-7,exp-8} has become one of the  center
 attentions in Particle Physics, long-baseline (LBL) experiments have played a significant role to
understand the hidden nature of the fundamental particle neutrino. With the confirmation of non-zero reactor mixing angle by both accelerator and reactor neutrino experiments, the three flavor neutrino oscillation paradigm which is governed by two mass
squared differences ($\Delta m_{21}^2$ and $\Delta m_{31}^2$ ) has become the most accepted theoretical model for the neutrino flavor transition.  The unitary mixing of active neutrinos ($\nu_e,\nu_{\mu}$, and $\nu_{\tau}$) in this model is described by three mixing angles ($\theta_{12}$, $\theta_{13}$, $\theta_{23}$) and one phase $\delta_{{\rm CP}}$. Though the oscillation parameters in this paradigm are determined with an unprecedented accuracy, the information about CP-violating phase, hierarchy of neutrino masses, and  octant of $\theta_{23}$ (i.e., whether $\theta_{23}$ is greater or lesser than $45^o$) are not known. The determination of these unknows by currently running long-baseline experiments (T2K and NO$\nu$A) is quite challenging as there exists degeneracies among the oscillation parameters and data collected so far are not sufficient enough to resolve the degeneracies among the parameters. The phase II runs of current generation experiments along with future generation LBL experiments with  greater energy resolution, improved statistics, and magnificent matter effect like Deep Underground Neutrino Experiment (DUNE), Tokai to Hyper-Kamioka (T2HK), Tokai to Hyper-Kamioka to Korea (T2HKK), and European Spallation Source Neutrino Super Beam (ESS$\nu$SB) etc, are expected to shed light on  remaining unknowns in neutrino sector.

The discovery of neutrino oscillation not only opens up a way to probe the properties of neutrino, but also  motivates to explore physics beyond Standard Model (SM) as it is clearly indicate that neutrino has non-zero tiny mass which is one of the shortcomings of SM. Consequently, many viable mechanisms collectively known as seesaw mechanisms \cite{Mohapatra:1979ia, Magg:1980ut, Schechter:1980gr, Lazarides:1980nt, Mohapatra:1980yp, Wetterich:1981bx, Ma:1998dn, Ma:2002pf, Hambye:2003rt, Mohapatra:1986bd} are introduced in the litterature to explain the lightness of neutrino mass. However, the  models based on low-scale seesaw \cite{Forero:2011pc} are more captivating over the high-scale seesaw as these models require strong evidence to support them and within the experimental limit it is difficult to probe high-scale seesaw even using Large Hadron Collider experiments. Whereas in low-scale seesaw models for instance, inverse seesaw \cite{Hirsch:2009mx, Schechter:1981cv, Ibarra:2003up1}, linear seesaw etc contain new neutrino states, which do not have any SM interaction, so-called sterile neutrinos with mass of the order of GeV/TeV scale as the the seesaw breaking scale in these models can be brought down to TeV/GeV scale. Therefore,  signatures of such sterile neutrinos can be probed at LHC experiments. Moreover, existence of such sterile neutrinos and their mixing with active neutrinos leads to the non-unitary mixing of active neutrinos. As a result, one can also probe  them at long baseline neutrino experiments by looking at the deviation from the unitary mixing of active neutrinos.

Enormous studies regarding the non-unitary neutrino mixing have been already discussed in the literature in both phenomenological and theoretical point of views\cite{Antusch:2006vwa,Goswami:2008mi, Dev:2009aw, Abada:2012mc, Abada:2013aba, Awasthi:2013ff, Abada:2014kba, Emelyanov:2014jna, Antusch:2016brq, Escrihuela:2015wra, Blennow:2016jkn}. In \cite{Malinsky:2009df}, it has been  shown that  non-unitary effects originated from a minimal inverse seesaw model can be probed at neutrino factory experiment. The bounds on non-unitary mixing parameters are obtained in \cite{Antusch:2008tz,Antusch:2014woa}. Moreover, the imapct of non-unitarity mixing  on the determination of  various unknowns in neutrino sector such as neutrino mass hierarchy, octant of atmospheric mixing angle, and CP violating phase by long-baseline experiments are discussed in \cite{Escrihuela:2016ube, Dutta:2016czj, Dutta:2016vcc, Verma:2016nfi}. A combined analysis of short and long-baseline neutrino oscillation data in non-unitary mixing scenario has been explored in \cite{Forero:2021azc} and it is found that there is no significant deviation from unitary mixing. The results of a combined analysis in neutrino oscillations without unitarity assumption in the three flavor mixing is presented in \cite{Hu:2020oba}. In \cite{Ellis:2020hus}, it is found  that with the next-generation  experimental data, the normalizations of all rows and columns of the lepton mixing matrix will be constrained to $\le$10\% precision, with the e-row best measured at $\le$1\% and the $\tau$-row worst measured at $\approx$10\% precision. A recent study on non-unitary mixing using  current generation experiments showed that the stronger tension which is existing between the latest 2020 data of the T2K and NO$\nu$A experiments gets reduced with non-unitary analysis \cite{Miranda:2019ynh}. Another study which obtained the constraints for non-unitarity coming from the observables: the neutrino-antineutrino gamma process and the invisible Z boson decay into neutrinos is presented in \cite{Escrihuela:2019mot}.   This paper address three  basic questions regarding the non-unitary mixing for the first time which  are
\begin{itemize}
\item Whether the long-baseline experiments based at J-PARC are capable of establishing the unitarity of active neutrino mixing matrix by ruling out non-unitary mixing or not?
\item What is the bounds on non-unitary mixing parameter that can be achieved by these experiments?
\item What is the sensitivity limits of these experiments in determining the non-unitary mixing parameter?
\end{itemize}
This paper is organised as follows. The neutrino oscillation in presence of non-unitary mixing scheme is discussed in Section II. Section III discusses the simulation details of the experiments which are considered for this study. The capability of these experiment in establishing unitarity of active neutrino mixing, the bounds on  non-unitary (NU) parameter and the sensitivity limits of these experiment in determining NU parameter are respectively discussed in Section IV. Finally, the summary and conclusions of this study is given in Section V.

\section{Neutrino oscillation in presence of non-unitarity mixing }
The general form of a unitary neutrino mixing matrix in a model with n sterile neutrino can be written as
 \begin{eqnarray}
 \mathcal{U} =\left( \begin{array}{cccccc}
 U_{e1}&U_{e2}& U_{e3} &. &. &U_{e(n+3)}\\
 U_{\mu 1}&U_{\mu 2}& U_{\mu 3} &. &. &U_{\mu (n+3)}\\
 U_{\tau 1}&U_{\tau 2}& U_{\tau 3} &. &. &U_{\tau (n+3)}\\
 U_{s^1 1} & U_{s^1 2} & U_{s^1 3}& . & . & U_{s^1 (n+3)}\\
 . & . & . & . & . &.\\
  . & . & . & . & . &.\\
  U_{s^n 1} & U_{s^n 2} & U_{s^n 3}& . & . & U_{s^n (n+3)}\\
 \end{array} \right) \approx 
 \left( \begin{array}{cc}
  N_{3\times 3}& \Theta_{3 \times n}\\
  R_{n\times 3} & S_{n \times n}
  \end{array} \right)  \text{    with    ~~~  }  \mathcal{U}\mathcal{U}^{\dagger}=I,
 \end{eqnarray}
where $N_{3\times 3}$ is the active neutrino mixing matrix which is no more unitary, $\Theta_{3 \times n}$ and $ R_{n\times 3}$ are active-sterile neutrino mixing matrix, and $S_{n \times n}$ is sterile-sterile neutrino mixing matrix. It should be also noted that the sub-matrix $\mathcal{W}$($=\left[N ~\Theta \right]$) of $\mathcal{U}$  satisfy the unitarity relation 
\begin{equation}
\mathcal{W}\mathcal{W}^{\dagger} = NN^{\dagger}+\Theta \Theta^{\dagger} = I
\end{equation}    Generally, the non-unitary active neutrino mixing matrix N is decomposed in two ways:
\begin{itemize}
\item[{i)}] in terms of $\eta ~(= \frac{1}{2}\Theta^{\dagger}\Theta)$ parameters,
\begin{equation}
N = (1-\eta) U_{PMNS}
\end{equation}
\item[{ii)}] in terms of lower triangular matrix T with parameter $\alpha$,
\begin{equation}
N=TU=(I-\alpha)U\;,\label{tm}
\end{equation}
where $U$ is the standard neutrino mixing matrix.The explicit form of triangular matrix T is given by
\begin{equation}
T=\left( \begin{array}{ccc}
\alpha_{11}&0&0\\
\alpha_{21}&\alpha_{22}&0\\
\alpha_{31}&\alpha_{32}&\alpha_{33}\\
\end{array}\right),
\end{equation}
where the diagonal elements of $T$ are of the form $ (1-\alpha_{ii}) \to \alpha_{ii} $.
\end{itemize}
 The relation between these two parametrizations of non-unitary mixing is derived in \cite{Blennow:2016jkn} and it is given by
\begin{equation}
\left( \begin{array}{ccc}
\eta_{11}&0&0\\
2\eta_{12}^*&\eta_{22}&0\\
2\eta_{13}^*&2\eta_{23}^*&\eta_{33}\\
\end{array}\right)= 
\left( \begin{array}{ccc}
\alpha_{11}&0&0\\
\alpha_{21}&\alpha_{22}&0\\
\alpha_{31}&\alpha_{32}&\alpha_{33}\\
\end{array}\right).
\end{equation}
This paper follows the parametrization of N in terms of triangular matrix as it is the preferred one for oscillation studies. In presence of non-unitary mixing the flavor state of neutrino can be written as
\begin{equation}
|\nu_{\alpha} \rangle = \sum_i N_{\alpha i} |\nu_i \rangle
\end{equation}
 As neutrino propagates the mass eigenstate evolves as
\begin{equation}
i\frac{d}{dt}|\nu_i \rangle = \mathcal{H}_0 |\nu_i \rangle\;,
\end{equation}
 where $\mathcal{H}_0$ is Hamiltonian in  vacuum, i.e.,
 \begin{eqnarray}
 \mathcal{H}_0 =\frac{1}{2E} \left( \begin{array}{ccc}
 0&0&0\\
 0&\Delta m_{21}^2&0\\
 0&0&\Delta m_{31}^2\\
 \end{array} \right)
 \end{eqnarray}
The non-unitary neutrino oscillation probability in vacuum is  given by
\begin{eqnarray}
P(\nu_{\mu}\to \nu_e) = \sum_{i,j}^3 N_{\mu i}^*N_{ei}N_{\mu j}N_{ej}^* &-&4 \sum_{j>i}^{3}Re\left[N_{\mu j}N_{ej}N_{\mu i}N_{ei}^*\right] \sin^2\left(\frac{\Delta m_{ji}^2L}{4E} \right)\nonumber \\
&+& 2 \sum_{j>i}^3 Im\left[N_{\mu j}^*N_{ej}N_{\mu i}N_{ei} \right] \sin \left(\frac{\Delta m_{ji}^2L}{2E} \right)
\end{eqnarray}
And the explicit form by neglecting cubic products of $\alpha_{21},\sin\theta_{13},$ and $\Delta m_{21}^2$ gives \cite{Escrihuela:2015wra}
\begin{equation}
P(\nu_{\mu}\to \nu_e) = (\alpha_{11}\alpha_{22})^2 P_{\mu e}^{SO}+\alpha_{11}^2\alpha_{22}|\alpha_{21}|P_{\mu e}^I+\alpha_{11}^2|\alpha_{21}|^2
\end{equation}
where $P_{\mu e}^{SO}$ is the vacuum neutrino oscillation probability in standard three flavor oscillation framework i.e.,
\begin{eqnarray}
P_{\mu e}^{SO} &=& \sin 2\theta_{12}\cos^2\theta_{23}\sin^2\left(\frac{\Delta m_{21}^2L}{4E} \right) + \sin 2\theta_{13}\sin^2\theta_{23}\sin^2\left(\frac{\Delta m_{31}^2L}{4E} \right) \nonumber \\
&&+ \sin2\theta_{12}\sin2\theta_{23}\sin\theta_{13}\sin\left( \frac{\Delta m_{21}^2L}{2E}\right) \sin\left(\frac{\Delta m_{31}^2L}{4E} \right)\cos\left(\frac{\Delta m_{31}^2L}{4E}-I_{123} \right),
\end{eqnarray}
and $P_{\mu e}^I$ is the term which contain the new phase and explicit form is given by
\begin{eqnarray}
P_{\mu e}^I&=& -2 \left[\sin2\theta_{13}\sin\theta_{23}\sin \left(\frac{\Delta m_{31}^2L}{4E}+I_{NP}-I_{123}\right) \right] \nonumber \\
&-& \cos\theta_{13}\cos\theta_{23}\sin 2\theta_{12}\sin\left( \frac{\Delta m_{21}^2L}{2E}\right)\sin(I_{NP}),
\end{eqnarray}
with $I_{123}=-\delta_{CP}=\phi_{12}-\phi_{13}+\phi_{23}$ and $I_{NP}=\phi_{12}-Arg(\alpha_{21})$. 
The propagation of neutrino through matter is governed by the charged current ($V_\text{CC}=\sqrt{2}G_F n_e$) and neutral current ($V_\text{NC}=-G_F n_n/\sqrt{2}$) matter potentials. In presence of non-unitary neutrino mixing, the CC and NC interaction Lagrangian becomes \cite{Antusch:2006vwa}
\begin{equation}
-\mathcal{L}_{int} =V_\text{CC} \sum_{i,j} N^*_{ei}N_{ej} \bar{\nu}_i \gamma^0 \nu_j +V_\text{NC}\sum_{\alpha,i,j}N^*_{\alpha i}N_{\alpha j}\bar{\nu}_i
\gamma^0\nu_j\;,
\end{equation} 
which yields the effective Hamiltonian as
\begin{eqnarray}
\mathcal{H}^N_m =\frac{1}{2E} \left( \begin{array}{ccc}
0&0&0\\
0&\Delta m_{21}^2&0\\
0&0&\Delta m_{31}^2\\
\end{array} \right)+ N^{\dagger} \left( \begin{array}{ccc}
V_\text{CC}+V_\text{NC}&0&0\\
0&V_\text{NC}&0\\
0&0&V_\text{NC}\\
\end{array} \right) N.
\end{eqnarray}
Then the non-unitary oscillation probability after travelling a distance L yeilds
\begin{equation}
P_{\alpha \beta} (E,L) =|\langle\nu_{\beta}|\nu_{\alpha}(L)|^2 = \left |\left(N e^{-i{\cal H}_m^N L}N^{\dagger}\right)_{\beta \alpha}\right |^2.
\end{equation}  
An attempt to  obtain the explicit analytical expression for neutrino oscillation probability in presence of non-unitary mixing is done in \cite{MY}, by using the formalism given in \cite{Fong:2017gke,TH}. Further, an explicit perturbative calculation up to the first order in the $\nu_{\mu} \rightarrow \nu_e$  oscillation channel  has been done in \cite{Martinez-Soler:2019noy,Fong:2017gke}. However, in this paper  numerical calculations  is done by using the General Long Baseline Experiment Simulator (GLoBES)~\cite{Huber:2004-1,Huber:2009-2} package along with the plugin MonteCUBES \cite{Blennow:2009}. The  neutrino oscillation parameters which are considered in this analysis are given in the Table \ref{oscl}.  From the previous analysis on nonunitary parameters \cite{C:2017scx,Soumya:2018nkw}, it is identified that the parameters in the 21 sector plays major role in $\nu_{\mu}$ to $\nu_e$ oscillation channel. Therefore, now onwards, the discussion is focused on non-unitary parameters in 21 sector i.e., $\alpha_{21}$ and its corresponding new phase $\phi_{21}$. 
\begin{table}
\begin{center}
\begin{tabular}{|c|c|c|c|c|c|c|c|} \hline
Parameters & $\sin^2\theta_{12}$& $\sin^2 2\theta_{13}$ & $\sin^2 \theta_{23}$ & $\Delta m_{21}^2$&$\Delta m_{atm}^2$ & $\delta_{CP}$ \\
& & & & & NH (IH) & \\ \hline
Best fit & 0.307 & 0.085 & 0.5 & $7.4 \times 10^{-5}~ {\rm eV}^2$ &$2.5 (-2.4) \times 10^{-3} ~{\rm eV}^2$ & $-90^\circ$\\
\hline
\end{tabular}
 \caption{{\label{oscl}}The values of neutrino oscillation parameters used in the analysis \cite{Esteban:2018azc}.}
\end{center}
\end{table}
Though the phase associated with the complex non-unitarity parameter can take values from $-\pi$ to $\pi$, the new phase is assumed  to be zero while doing the analysis unless otherwise mentioned.

\begin{figure}
\begin{center}
\includegraphics[scale=0.5]{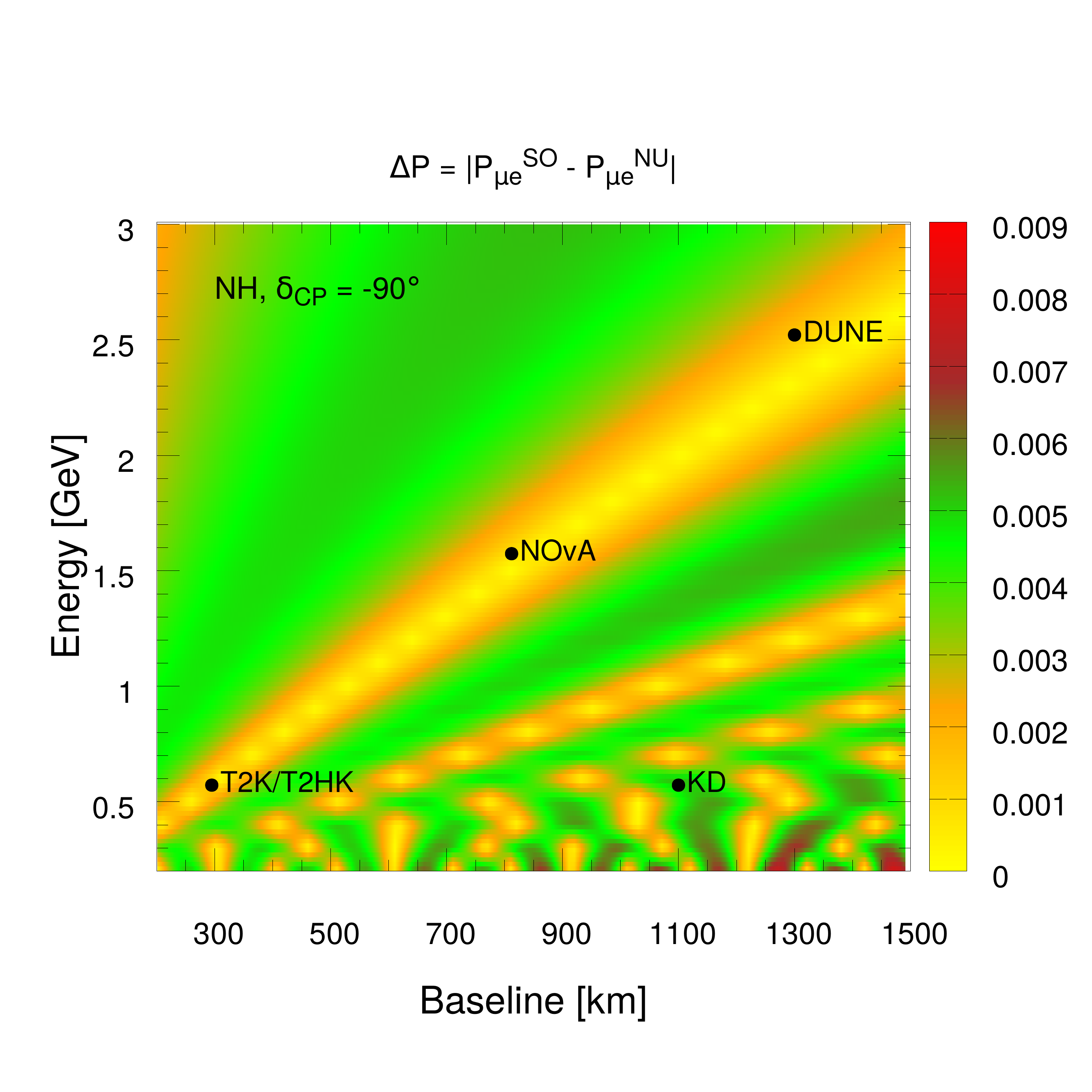}
\end{center}
\caption{{\label{newle}} $\Delta P_{\mu e}$ in $L-E$ plane for the non-unitarity parameter $\alpha_{21}=0.01$. The  mass hierarchy of neutrino is assumed to be normal and other oscillation parameters are taken as given in Table \ref{oscl}.}
\end{figure} 

To quantify the deviation from  unitary mixing of neutrinos, one can define  $\Delta P_{\mu e} = \displaystyle{|P_{\mu e}^{NU}-P_{\mu e}^{SO}|}$, where $P_{\mu e}^{NU}$ and  $P_{\mu e}^{SO}$ respectively are the oscillation probability in non-unitary and unitary mixing scheme. As the LBL experiments are mainly searching for $\nu_{\mu} \to \nu_{e}$  and $\bar{\nu}_{\mu} \to \bar{\nu}_{e}$  oscillation signals, the relative deviation in the $\nu_{\mu} \to \nu_{e}$ oscillation probability due to  the  non-unitary mixing as a function of all possible neutrino energies and baselines is shown in Fig.\ref{newle}. The color gradient  corresponds to relative deviation in oscillation probability.  From the figure, it can be seen that the non-unitary parameters in 21 sector can be probed at long-baseline experiments like T2K, NO$\nu$A, T2HK, T2HK, and DUNE.

\section{Simulation details}

This section briefly describes the experimental features of the LBL experiments at J-PARC:Tokai to Kamioka (T2K),  Tokai to Hyper-Kamioka (T2HK), and Tokai to Hyper-Kamioka to Korea (T2HKK) which are  considered in this analysis.

T2K \cite{Itow:2001ee, Ishitsuka:2005qi} is a currently running LBL experiment which has already started collecting  data from 2010 onwards, whereas T2HK and T2HKK are the proposed experiments which are considered as the upgraded version of T2K experiment. T2K experiment completed its scheduled run and now it is upgraded to Phase II  and continuing to take neutrino data. The muon neutrino/antineutrino beam for all these experiments is produced in the J-PARC accelerator facility at Tokai. However, the Water-Cerenkov detectors of  these experiments are located at different locations. The detectors of T2K and T2HK experiments are kept at 295km away from the neutrino beam source. T2HKK experiment has two detectors: The first detector so-called Japan Detector (JD) which is planned to keep at 295km away from the source  at Japan, and the second detector so-called Korean Detector (KD) is planned to keep at Korea about 1100km away from the source. Moreover, the fiducial mass of detectors of each experiments are different. The detector fiducial mass of T2K experiment is 22.5kt. Initial plan of T2HK experiment is to consider 560kt fiducial mass for the detector. However, the recent plan of this experiment is to consider 374kt. Each of two detectors of T2HKK experiment (JD and KD) is having a fiducial mass of 187kt. The detector of T2HK is also known as 2JD as the fiducial mass of T2HK detector is twice that of JD detector of T2HKK. The detector of all these experiments is kept at an off-axis angle 2.5$^{\circ}$ to the neutrino beam line which helps the neutrino flux to peak sharply at first (second) oscillation maximum of 0.6 GeV for detector which is kept at Japan (Korea). Further, such off-axis beam nature also  reduces the intrinsic $\nu_e$ contamination in the beam and the background due to neutral current events and thus helps to improve the signal-to-background ratio by great extent. \\
A proton beam power of 750 kW and with a proton energy of 30 GeV which corresponds to a total exposure of 7.8$\times 10^{21}$ protons on target (POT) with 1:1 ratio of neutrino to antineutrino modes is considered to simulate T2K experiment for this study\cite{Abe:2017uxa}. The signal and background event spectra and rates are matched with that given in the recent publication of the T2K collaboration \cite{Abe:2014tzr}. An uncorrelated 5\% normalization error on signal and 10\% normalization error on background for both the appearance and
disappearance channels  are considered as given in \cite{Abe:2014tzr} to analyze the prospective data from the T2K experiment and assumed  that the set of systematics for both the neutrino and antineutrino channels are uncorrelated.\\
A total of 10 years of operation with 1.3 MW beam power with 1:3 ratio of neutrino to antineutrino modes which corresponds to 27$\times 10^{21}$ proton on target (POT) by following \cite{Abe:2016ero,Abe:2018uyc} is considered to simulate both T2HK and T2HKK experiments. Moreover, both signal and background event spectra and rates are matched with those  given in \cite{Abe:2018uyc}. An uncorrelated 5\% normalization error on signal and 10\% normalization error on background for both the appearance and disappearance channels are used as the way it is considered those for T2K experiment. For the simulations, GLoBES along with MonteCUBES have been used. Further, the Poissonian  $\chi^{2}$ is evaluated using GLoBES package \cite{MCG,Fogli_2002,Huber:2002mx} and its explicit from is given by
 \begin{eqnarray} \label{cchi}
 \chi^{2} =\underset{\xi_l,\vec{p}}{\mathrm{ min}} \sum_{i=1}^{N_E} \left[ 2 N_i^{\mathrm{th}}\left( \vec{P},\xi\right)-2 N_i^{\mathrm{dat}} \left(\vec{P},\xi\right)-2 N_i^{\mathrm{dat}}\left( \vec{P},\xi\right)\mathrm{ln}\left( \frac{N_i^{\mathrm{th}}\left( \vec{P},\xi\right)}{N_i^{\mathrm{dat}}\left( \vec{P},\xi\right)}\right)\right]+\sum_{l=1}^{2} \xi_l^2,\nonumber \\
\end{eqnarray}
where $N_i^{\mathrm{th}}\left( \vec{P},\xi\right)$ and $N_i^{\mathrm{dat}}\left(\vec{P},\xi\right)$  respectively are the expected and observed  events (both signal and background) for a considered i-th energy bin. Further, $N_i^{\mathrm{th}}\left(\vec{P},\xi\right) =N_i^0\left(1+\sum_{l=1}^2 \pi_i^l \xi_l^2 \right)$ with $N_i^0$ as the number of events without systematics, $N_E$ is the total number of energy bins, $\xi_1$ and $\xi_2$ are the systematic errors associated with signal and background events respectively, $\vec{P}=\left\lbrace \theta_{12},\theta_{13},\theta_{23},\Delta \mathrm{m}^{2}_{21}, \Delta \mathrm{m}^{2}_{31},\delta_{\mathrm{CP}}\right\rbrace$ representing all the fundamental oscillation parameters, while  $\vec{p}=\{\theta_{23},~\delta_{\mathrm{CP}},~ \Delta m_{31}^2\}$ is the subset of  $\vec{P}$ on which we perform marginalization. The marginalization range for $\delta_{\rm{CP}}$, $\sin^2\theta_{23}$, and $\Delta m_{31}^2$  are $[-180^{o}:180^{o}]$, $[0.4:0.6]$, and $[0.36:0.64]$ respectively. Moreover, in this work,  $\Delta \chi^{2}$  is determined using the pull variable over the systematic uncertainties  and a detailed discussion on this is given in \cite{Fogli_2002,Huber:2002mx}.


\section{Results and Discussions}
This section mainly  discuss the cabability of  LBL experiments based at J-PARC in establishing unitarity of neutrino mixing matrix. Further, this section also discuss the bounds on NU parameter and  sensitivity limits of LBL experiment to determine NU parameter.
\begin{figure}[!htb]
\begin{center}
\minipage{0.5\textwidth}
    \includegraphics[scale=0.6]{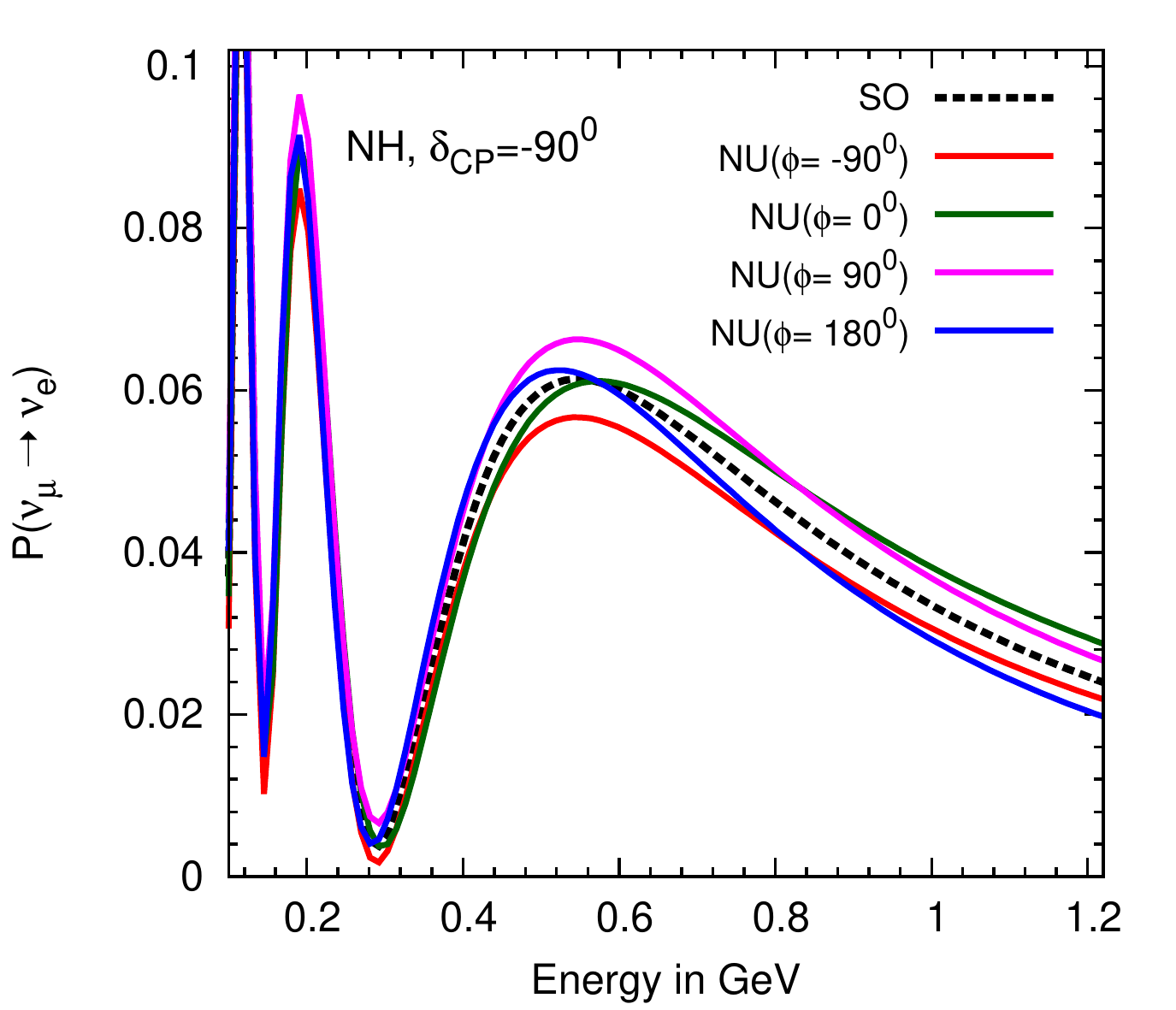} 
    \endminipage\hfill
    \minipage{0.5\textwidth}
    \includegraphics[scale=0.6]{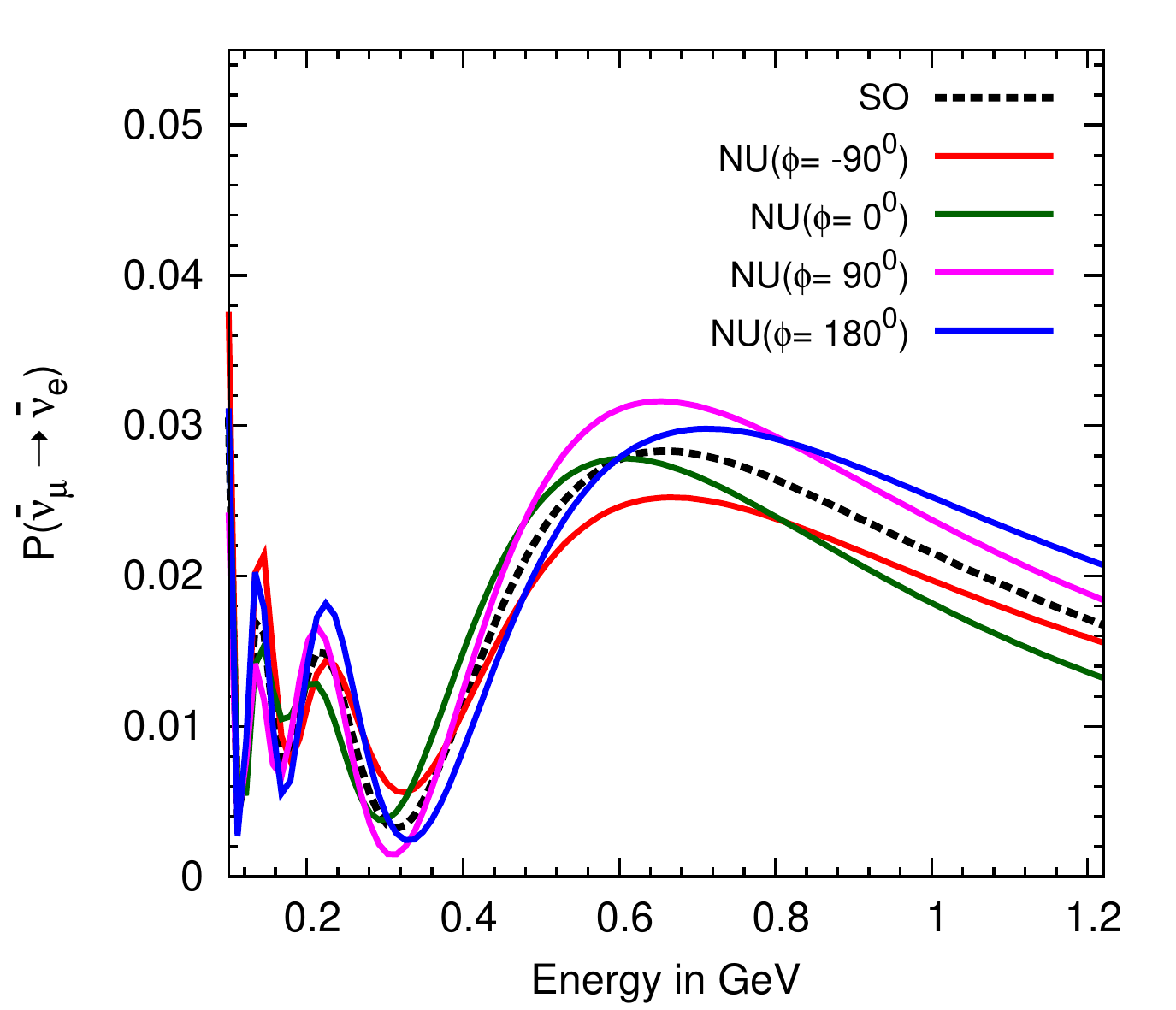} 
    \endminipage\hfill
  \end{center}
\caption{{\label{nuosc}} The black dashed curve corresponds to the oscillation probability in standard oscillation paradigm with $\delta_{CP}=-90^0$, whereas the red, green, magenta, and blue solid curves  correspond to oscillation probabilities in the presence of NU parameter $\alpha_{21}=0.01$ with $\phi_{21}=-90^0, 0^0, 90^0,$ and $180^0$ respectively. Neutrino mass hierarchy is assumed to be normal. The left (right) panel is for oscillation probability for neutrino (antineutrino). }
\end{figure} 

As this study is focusing on the non-unitarity parameters $\alpha_{21}$ and its corresponding CP-violating phases, it is most important to know how these parameters affect the oscillation probability. Though it is always better to start  with analysis by looking at the analytical expression of neutrino osillation probability to have a deep understanding of physics,  oscillation probability is calculated  numerically as a function of neutrino energy for this study and it is shown in Fig. \ref{nuosc}. In the figure, the black dashed curve corresponds to the oscillation probability in standard oscillation case with normal hierarchy and $\delta_{CP}=-90^0$, whereas the red, green, magenta, and blue solid curves  correspond to oscillation probabilities in the presence of NU parameter $\alpha_{21}=0.01$ with $\phi_{21}=-90^0, 0^0, 90^0,$ and $180^0$ respectively. The left (right) panel is for oscillation probability for neutrino (antineutrino). 
From the figure, it can be seen that for $\phi_{21}=90^0, -90^0$, there is a significant deviation from standard oscillation case, whereas there is no significant deviation for $\phi_{21}=0^0, 180^0$. Therefore, $\phi_{21}=90^0, -90^0$ are the favourable values of new phase to rule out the nonunitary mixing and $\phi_{21}=0^0, 180^0$ are the unfavourable values as there is no much deviation from standard oscillation case. It should be also noted that the oscillation curves for $\phi_{21}=0^0, 180^0$ is touching the standard oscillation curve at some point where one cannot distinguish between non-unitary mixing and unitary mixing. However, such intersection point is different in neutrino and antineutrino oscillation channels. Therefore, an interplay of neutrino and antineutrino oscillation helps in distinguishing unitary mixing from non-unitary mixing. 

\begin{figure}
\begin{center}
\minipage{0.5\textwidth}
    \includegraphics[scale=0.6]{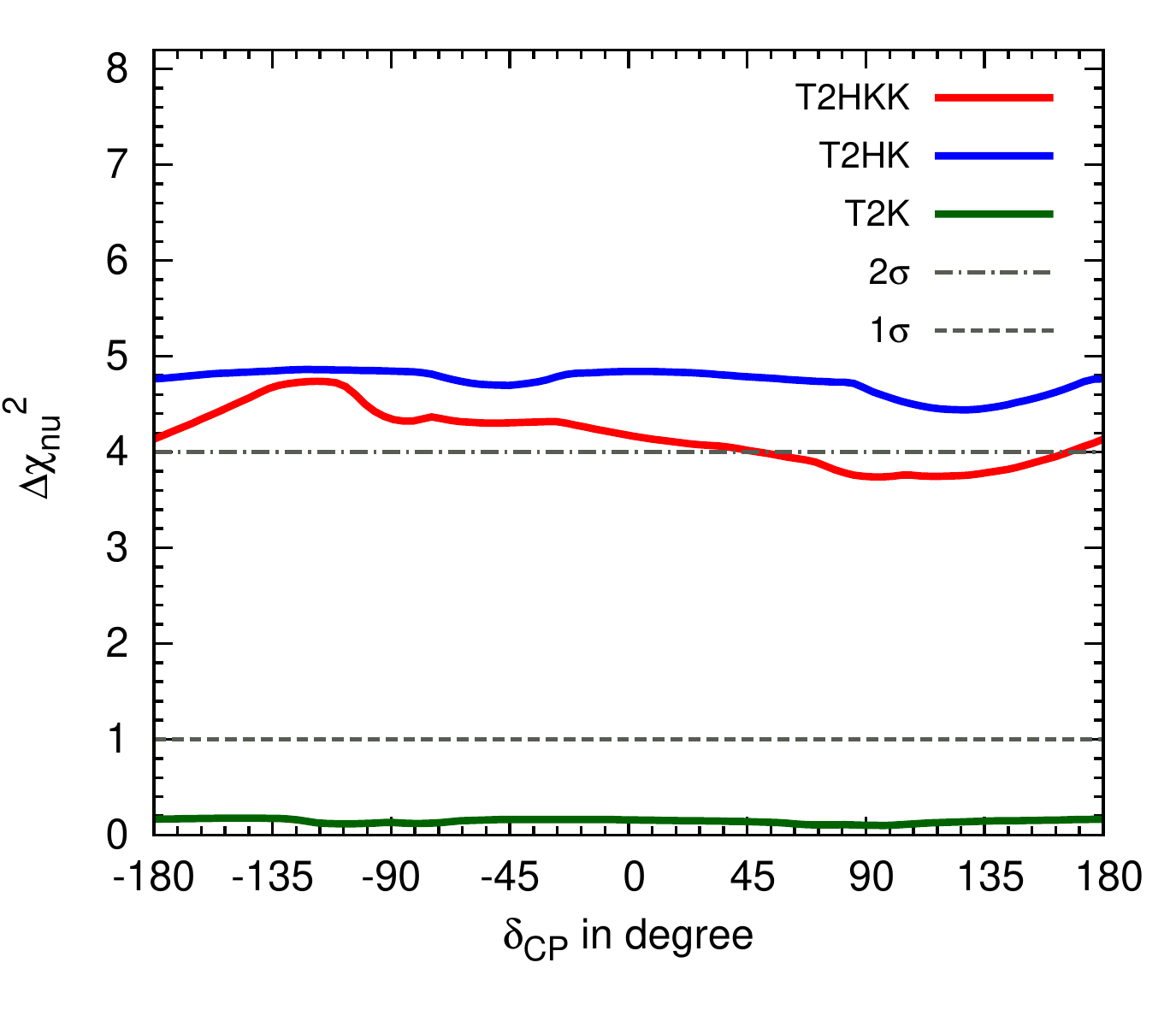}
 \endminipage\hfill
\minipage{0.5\textwidth}
     \includegraphics[scale=0.6]{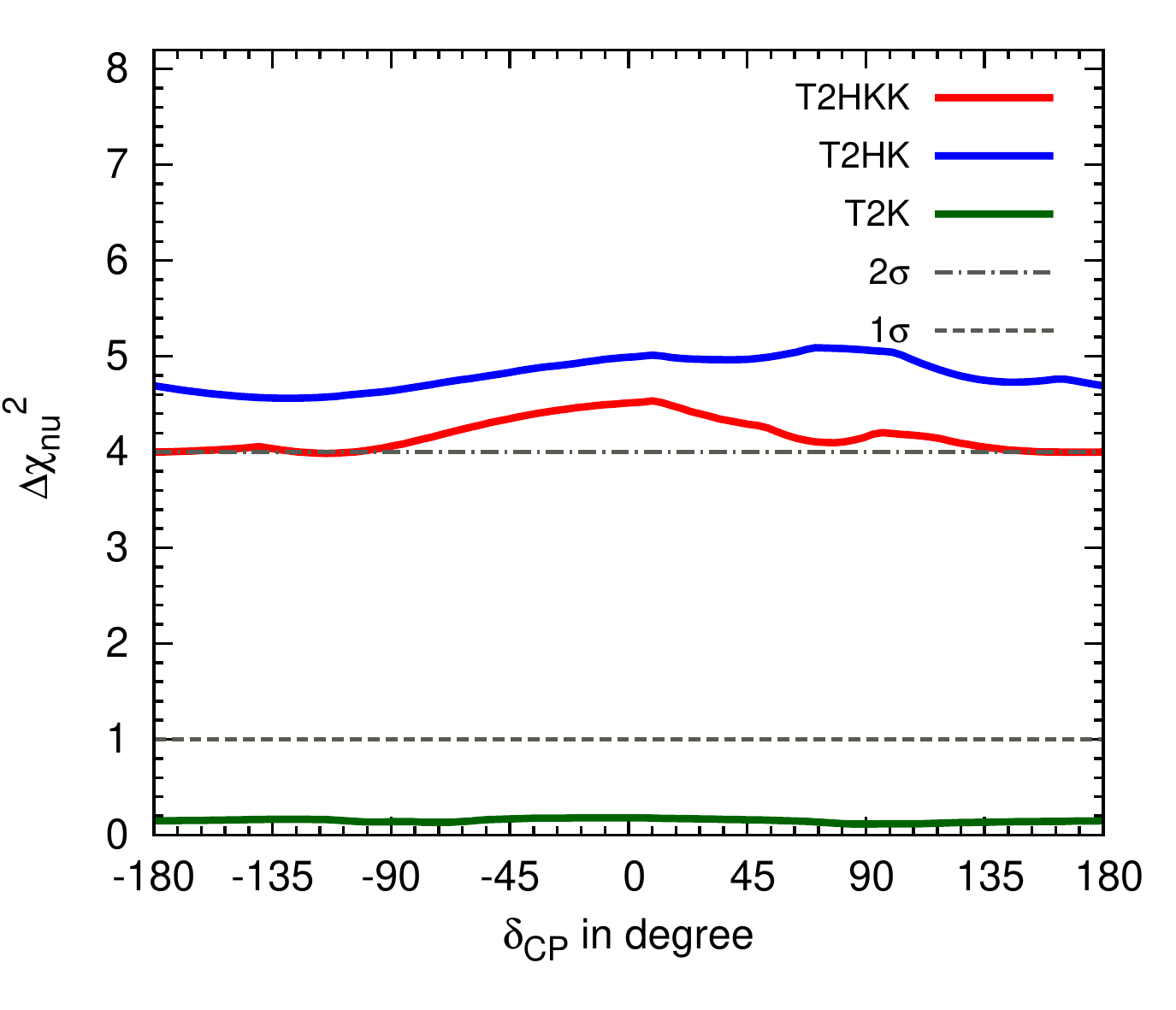}
   \endminipage\hfill
  \end{center}
\caption{{\label{chi}} The capability of LBL experiments in establishing the unitary mixing by ruling out the non-unitary mixing as a function of true values of $\delta_{CP}$. In the left (right) panel the  mass hierarchy of neutrino is assumed to be normal (inverted).}
\end{figure}

Now the focus will be on the unfavourable values of new phase, i.e,.$\phi=0^0$ as it is challenging for this value to distinguish non-unitary mixing from unitary mixing. In order to show the capability of LBL experiment to establish unitary mixing as a function of true values of $\delta_{CP}$, one can define
\begin{equation}
\Delta \chi^2_{NU} = \chi^2_{SO}-\chi^2_{NU},
\end{equation} 
where $\chi^2_{SO}$ is evaluated using Eqn.\ref{cchi} by assuming both $N^{\mathrm{th}}$ and $N^{\mathrm{dat}}$ are with unitary mixing. Whereas, $\chi^2_{NU}$ is calculated by assuming $N^{\mathrm{th}}$ with non-unitary mixing and $N^{\mathrm{dat}}$ with unitary mixing. The minimum value for $\Delta \chi^2_{NU}$ is obtained by doing marginalization over oscillation paramters including new phase $\phi_{21}$ in its allowed range $[-180^0:180^0]$. The minimized  $\Delta \chi^2_{NU}$ as a function of true values of $\delta_{CP}$ is shown in Fig.\ref{chi}. In the left (right) panel of the figure hierarchy is assumed to be normal (inverted) and the atmospheric mixing angle is set to maximal mixing. From the figure, it can be seen that the T2K experiment can not rule out the non-unitary mixing with the data so far collected. Whereas  T2HKK can rule out non-unitary mixing  with a significance more than 2$\sigma$ for most of the values of $\delta_{CP}$. Moreover, T2HK can rule out non-unitary mixing above  2$\sigma$ C.L.  irrespective of mass hierarchy and the true value of $\delta_{CP}$. From the analysis, it is found that the interplay between neutrino and antineutrino runs helps in ruling out the non-unitary mixing.

 \begin{figure}
\begin{center}
\minipage{0.5\textwidth}
    \includegraphics[scale=0.5]{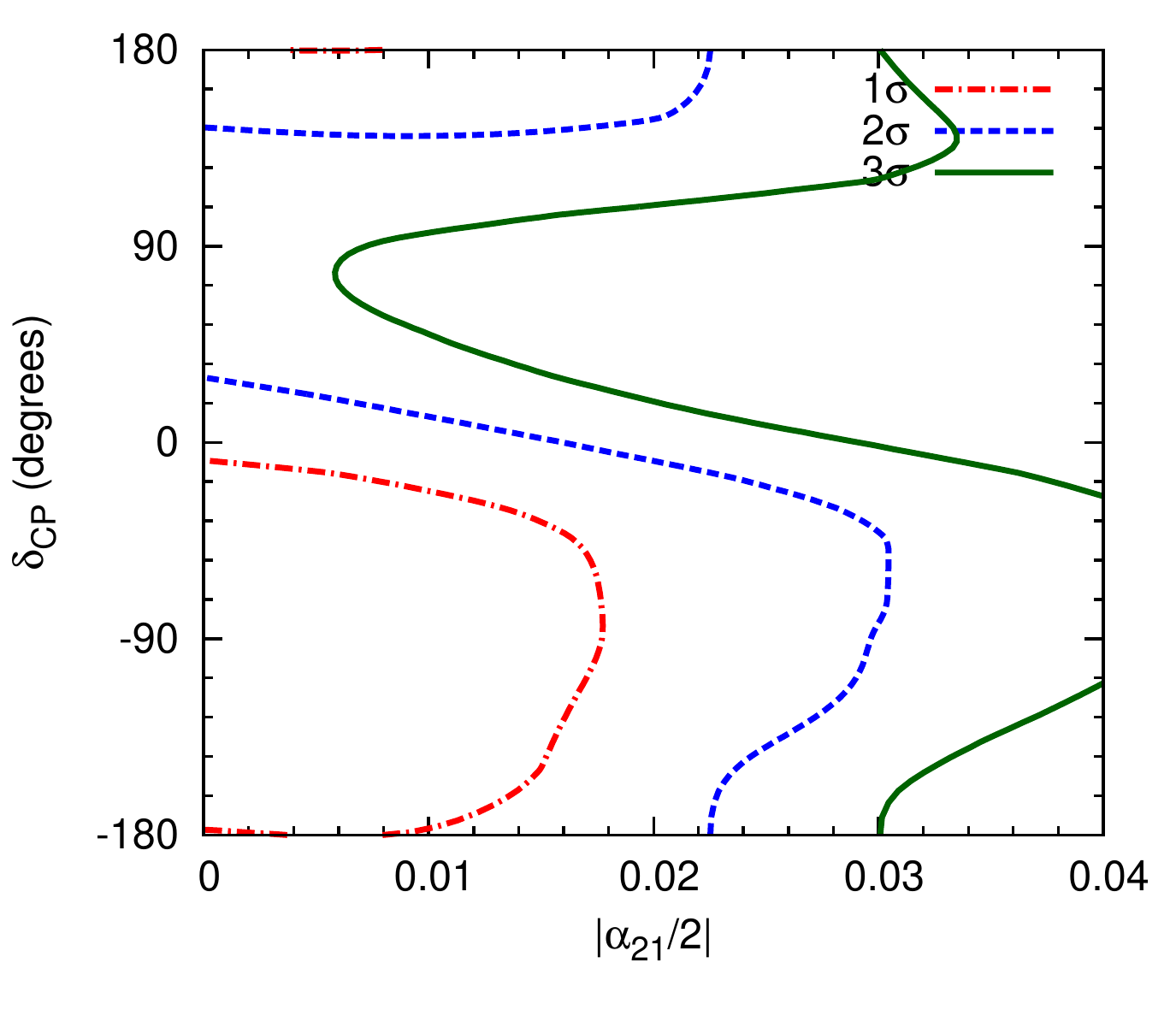} 
    \endminipage\hfill
    \minipage{0.5\textwidth}
    \includegraphics[scale=0.5]{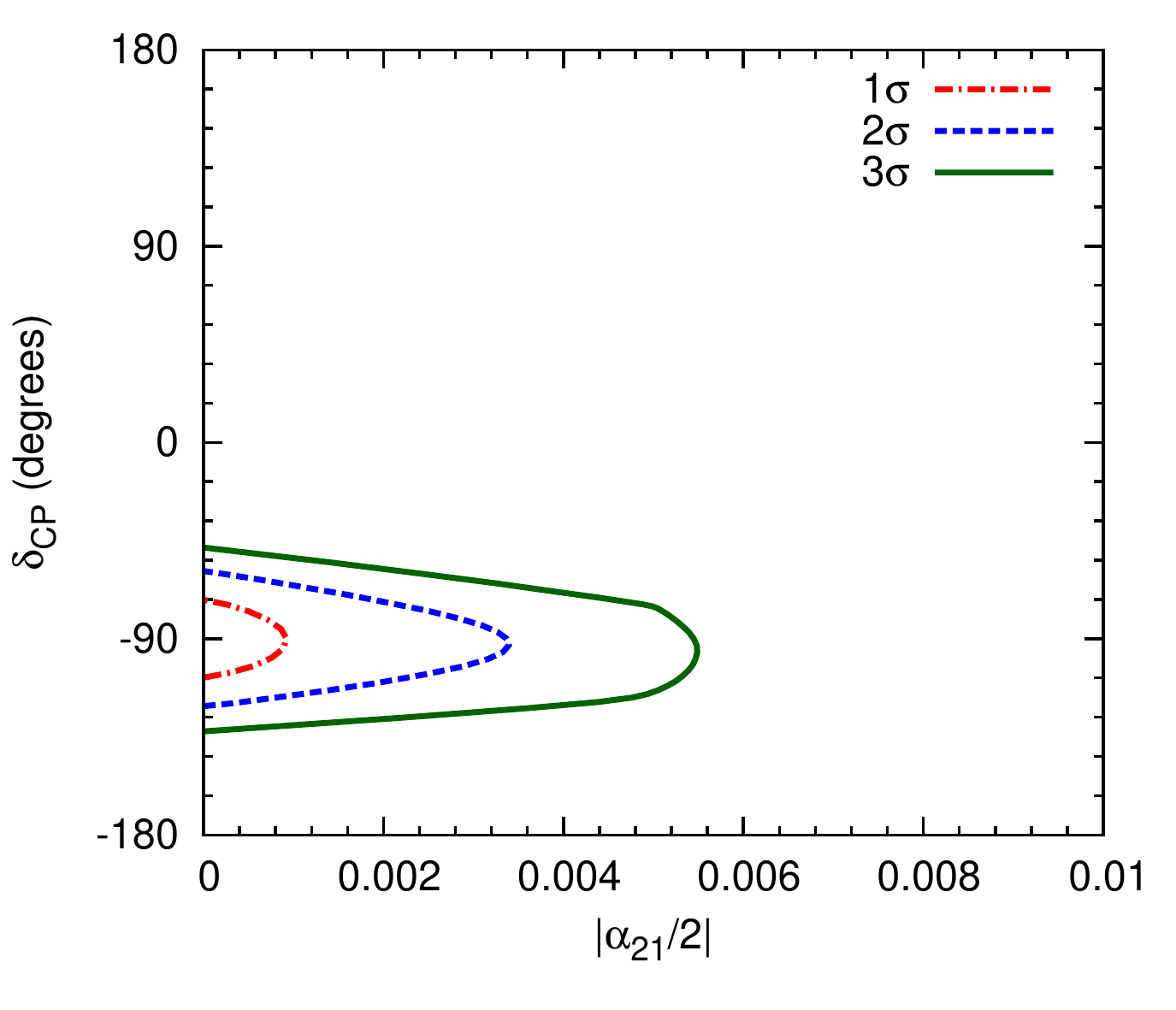} \\
    \endminipage\hfill
   \minipage{0.5\textwidth}
    \includegraphics[scale=0.5]{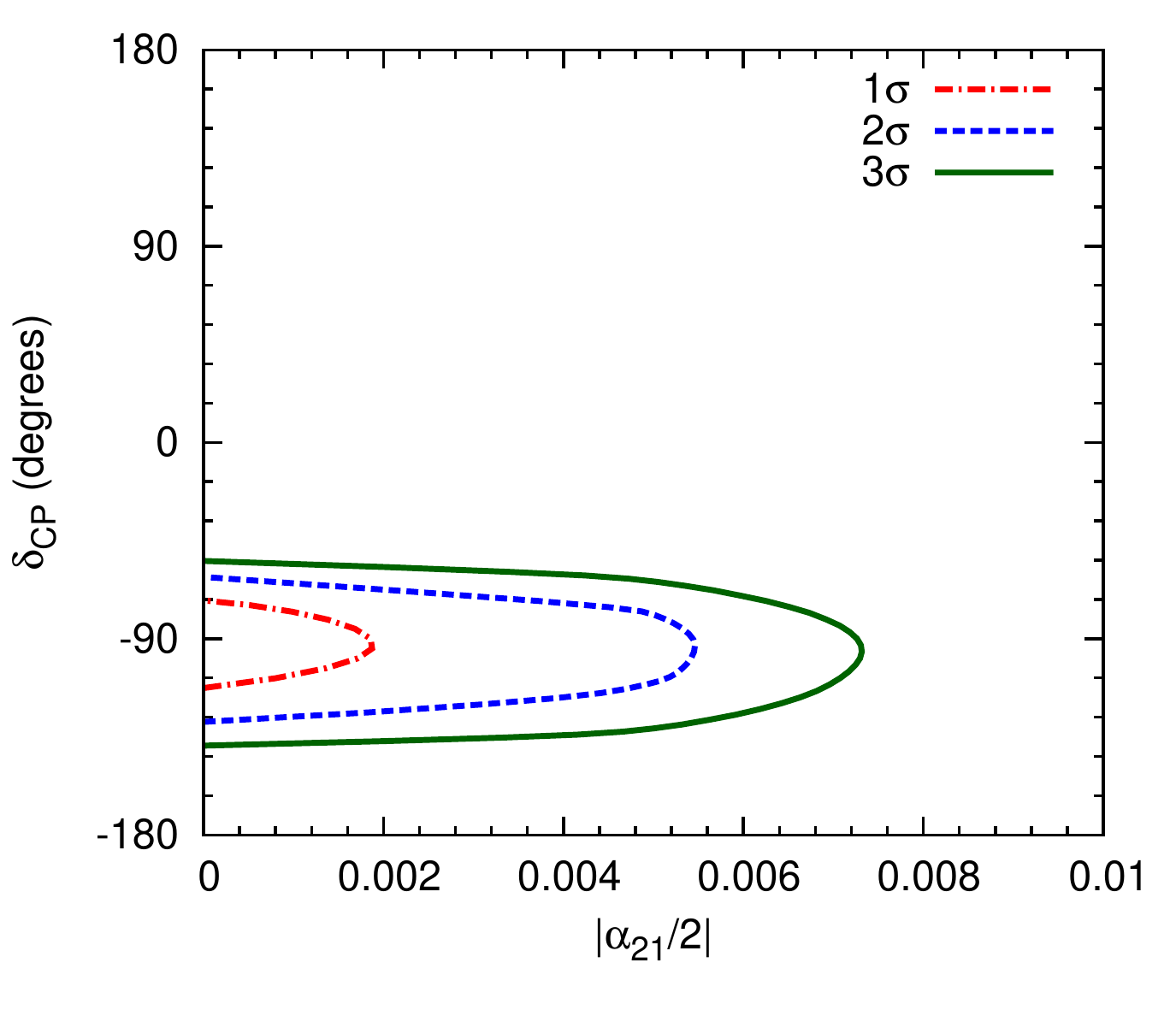}\\
     \endminipage\hfill
  \end{center}
\caption{{\label{bound}} The bounds on NU parameter is obtained by comparing unitary mixing against non-unitary mixing. The upper left (right) panel is for T2K (T2HK) and lower panel is for T2HKK. The red, blue, and green curves are respectively for 1$\sigma$, 2$\sigma$, and 3$\sigma$ C.L. contours.The neutrino mass hierarchy is asuumed to be normal and other oscillation parameters are used as given in Table \ref{oscl}.}
\end{figure} 

The bounds on NU parameters using LBL experiment is obtained by  comparing unitary mixing with oscillation parameters as shown in Table \ref{oscl} against non-unitary mixing. The minimised $\Delta\chi^2$ is evaluated by doing marginalization over oscillation parameters and it is shown in $\delta_{CP}^{test}-\left(\alpha_{21}/2\right)^{test}$ plane as given  in Fig.\ref{bound}. The red, blue, and green are respectively 1$\sigma$, 2$\sigma$, and 3$\sigma$ C.L. contours. From the figure, it can be seen that the bounds from T2K experiment on NU parameter is not  significantly constraint, whereas that for T2HK and T2HKK is  severly constraint. Further, the bounds on $\alpha_{21}/2$ are 0.028, 0.0026, 0.005 at 2$\sigma$ C.L. respectively for T2K, T2HK, and T2HKK.

\begin{figure}
\begin{center}
\includegraphics[scale=0.6]{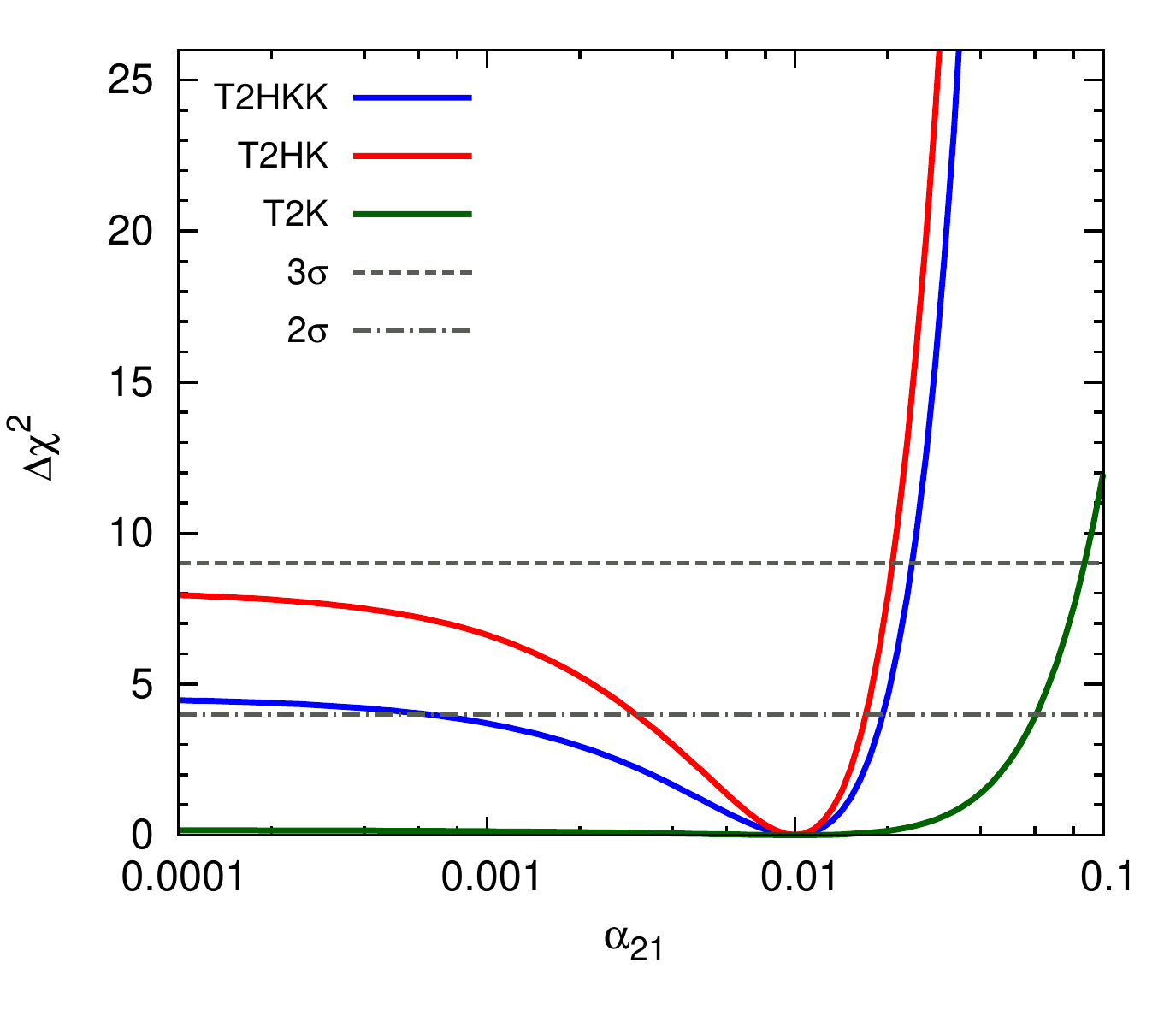}\\
\end{center}
\caption{{\label{pre}} The precision measurement of $\alpha_{21}$ at LBL experiments. The  mass hierarchy of neutrino is assumed to be normal and other oscillation parameters are set to as given in Table \ref{oscl}.}
\end{figure}

 Next, analysing the sensitivity limit of LBL experiments in determining NU parameters. In order to obtain this, the experimental data is simulated by fixing true oscillation parameters as given in Table \ref{oscl} and setting true value of $\alpha_{21}= 0.01$ , then comparing it with theory by varying $\alpha_{21}$ in the range [0.001:0.1]. Further, the  $\Delta\chi^2$ is minimised by doing marginalization over oscillation parameters and non-unitary phase and it is given in Fig.\ref{pre}. From the figure, it can be seen that the precision in the measurement of $\alpha_{21}$ by T2HK is better than that of both T2K and T2HKK.
 
\section{Summary and Conclusions}
The measurement of neutrio oscillation parameters in a three flavor framework is usually done by assuming that the active neutrino mixing matrix is unitary. Howbeit, the extended theories to accomodate massive neutrinos indicate the existence of new neutrino states which can give rise to non-unitary mixing of active neutrinos. This paper mainly scruitinized  whether  the long baseline experiments like T2K, T2HK, and T2HKK can establish the unitarity of active neutrino matrix by ruling out such non-unitary mixing in 21 sector. It is found that T2HK can establish unitarity of active neutrino mixing above 2$\sigma$ C.L. irrespective of neutrino mass hierarchy and true value of $\delta_{CP}$ if NU parameter $\alpha_{21}$ of the order of $10^{-2}$. Further, this paper is also investigated  the bound on NU parameter that can be acieved from these LBL experiments and found that the bounds on $\alpha_{21}/2$ are 0.028, 0.0026, 0.005 at 2$\sigma$ C.L. respectively for T2K, T2HK, and T2HKK. Finally, it is also found that the sensitivity limit of T2HK on NU parameter is far better than that of both T2HKK and T2K.
\vspace{1cm}\\
{\bf Acknowledgments:}\\
Author would like to thank the organizers of the workshop XIX International Workshop on Neutrino Telescopes held online during 18-26 February, 2021 for giving an opportunity to present the preliminary results of this work. Author also would like to thank the Science and Engineering Research Board (SERB) for the funding under National Post-Doctoral Fellowship (NPDF) scheme [PDF/2019/003346].

\end{document}